# Astrophysical *S*-factor for the radiative capture reaction p$^{13}$C→$^{14}$Nγ

S. B. Dubovichenko,[*] A. V. Dzhazairov-Kakhramanov[†]

*V.G. Fessenkov Astrophysical Institute "NCSRT" NSA RK, Almaty, Kazakhstan*

N. A. Burkova[‡]

*Al-Farabi Kazakh National University, Almaty, Kazakhstan*

The phase shift analysis, done on the basis of the known measurements of the differential cross-sections of the p$^{13}$C elastic scattering at the energy range 250÷750 keV, shows that it is enough to take into account only $^3S_1$ wave in the considered energy region. The potential for the triplet $^3S_1$ state in p$^{13}$C system at the resonance energy 0.55 MeV corresponding to quantum numbers $J^{\pi}T = 1^-1$ as well as the potential for the $^3P_1$ bound state of $^{14}$N were constructed on the basis of the obtained scattering phase shifts. The possibility to describe the experimental data of the astrophysical *S*-factor of the p$^{13}$C radiative capture at the energies 0.03÷0.8 MeV was considered within the potential cluster model with the forbidden states. It was shown that we properly succeed in explanation of the energy behavior of the astrophysical *S*-factor for the p$^{13}$C radiative capture at the resonance energy range 0.55 MeV (laboratory system).

## I. INTRODUCTION

The astrophysical *S*-factor, which determines the reaction cross-section, is the main characteristic of any thermonuclear reaction in the Sun and the stars of our Universe, i.e. it defines the probability of the reaction behavior at very low energies [1]. The astrophysical *S*-factor may be obtained experimentally, but for the majority of interacting light nuclei, which take part in the thermonuclear processes it can be done in the energy range 100 keV÷1 MeV only; and the errors of its measurement may reach one hundred and more percents, more often than not [2]. But, its values with the minimal errors, are required for the real astrophysical calculations, for example, for developing and refinement of the evolution model of the stars [3] and our Universe [4,5] in the energy range from 0.1 to 100 keV, relevant to the temperatures in the star centre – about $10^6$ K÷$10^9$ K.

One of the methods for obtaining the astrophysical *S*-factor at zero energy, i.e. at the energy of the order of the 1 keV and less, is the extrapolation from the energies where it can be determined experimentally to the lower energies. It is the common way, which is used, first of all, after carrying out the experimental measurements of the cross-section of a certain thermonuclear reaction. However, essential experimental errors in the *S*-factor measurement lead to the big ambiguities in the carrying out of the extrapolations and this fact considerably reduces the value of such results. The up-to-date review for the different results of the main thermonuclear reactions in the pp-chain and the CNO cycle, for selected experimental data of the astrophysical *S*-factors and for certain methods of their calculations, was recently published in review [6].

The second and evidently most preferable way consists in the theoretical calculations of *S*-factor for a thermonuclear reaction, for example, the radiative capture process on the basis of a certain nuclear model [1]. This method is based on the quite obvious assumption that if such a model describes the experimental data for the astrophysical *S*-factor correctly in the energy region where these data exist, for example within 100 keV÷1 MeV, then it is reasonable to assume that this model will describe the form of the *S*-factor properly, at more low energies (about 1 keV) where the direct experimental measurements are impossible up today [7].

---

[*] Electronic address: dubovichenko@mail.ru
[†] Electronic address: albert-j@yandex.ru
[‡] Electronic address: natali.burkova@gmail.com

## II. MODEL AND METHODS

Generally, for the calculations of astrophysical *S*-factors we use the potential cluster model (PCM) of light atomic nuclei with the classification by the orbital states according to the Young schemes [8,9,10]. This model contains, in some cases, forbidden by the Pauli principle states (FS) in pair intercluster interactions [9]. The model gives a lot of relatively simple possibilities for the calculation of the different astrophysical characteristics, for example, the astrophysical *S*-factor of the radiative capture for electromagnetic transitions from the scattering states of clusters to the bound states (BS) of light atomic nuclei in cluster channels [7,11]. The choice of the potential cluster model is caused by the fact, that in lots of atomic nuclei the probability of forming of the nucleon associations (clusters), and the degree of their isolation from each other are comparatively high. This is confirmed by the numerous experimental data and different theoretical calculations implemented over the past fifty-sixty years [9].

The results of the phase shift analysis of the experimental data on the differential cross-sections of elastic scattering of corresponding free nuclei are used for the construction of phenomenological intercluster interaction potentials for the scattering states in the PCM [7]. The intercluster interaction potentials in the framework of the formal two-body scattering problem are chosen under condition of the best description of the obtained elastic scattering phase shifts [12,13]. In this case, for any nucleon system, the many-body character of the problem is taken into account by separation of single-body levels of this potential onto allowed (AS) and Pauli forbidden states [7,9].

However, the results of the phase shift analysis usually existing only in a restricted energy range, does not allow us to reconstruct unambiguously the interaction potential. Therefore, the requirement based on the reproduction of the bound energy of the nucleus in the corresponding cluster channel and the description of some other static nuclear characteristics is the additional requirement for construction of the intercluster potential of the BS. Thus, for the bound states of clusters, the potentials are constructed primarily in agreement with the requirement of description of the main characteristics of this state. This additional requirement is, certainly, the idealization of the real existed situation in the nucleus, since it supposes that there is 100% clusterization in the ground state. Therefore, the success of this potential model, which describes the system of *A* nucleons in the bound state, is determined by the value of real clusterization of $A_1 + A_2$ nucleons in the BS channel of this nucleus. Nevertheless, some nuclear characteristics of certain, not cluster, nuclei can mainly be caused by the specified cluster channel, i.e., have the specified cluster structure with the small contribution of the other possible cluster configurations. In this case, the using one-channel cluster model allows one to identify the dominated cluster channel, also select and describe those characteristics of the nuclear systems, which are caused by such configuration [7].

Let us consider the classification of the orbital states of the p$^{13}$C system according to the Young's schemes. It was shown earlier that the Young's scheme {4441} corresponds to the ground bound state (GS) of $^{13}$N so as for $^{13}$C [8,14]. Let us remind that the possible orbital Young's schemes in the $N = n_1 + n_2$ system of particles can be characterized as the direct outer product of the orbital schemes of each subsystem, and, for the p$^{13}$C system within the 1*p*-shell, it yields $\{1\} \times \{4441\} \rightarrow \{5441\} + \{4442\}$ [9,15]. The first of the obtained scheme is compatible with the orbital moment $L = 1$ and is forbidden, so far as it could not be five nucleons in the *s*-shell, and the second scheme is allowed and is compatible with the orbital moments $L = 0$ and 2 [9,15]. Thus, in the $^3S_1$-potential there is the only one allowed state and the $^3P$-wave has the forbidden state and allowed one at the energy -7.55063 MeV [16]. However, the above-obtained result should be considered only as the qualitative estimation of possible orbital symmetries in the GS of $^{14}$N for the p$^{13}$C channel.

The standard expression [20] was used for the calculations of the astrophysical *S*-factor:

$$S(NJ, J_f) = \sigma(NJ, J_f) E_{cm} \exp\left( \frac{31.335 Z_1 Z_2 \sqrt{\mu}}{\sqrt{E_{cm}}} \right), \quad (1)$$

where σ is the total cross-section of the radiative capture process in barn, its theoretical expressions are given [7,21,22]; $E_{cm}$ is the particle center mass energy in keV; μ is the reduced mass in the entrance channel in atomic mass units (amu); $Z_{1,2}$ are the particle charges in units of elementary charge; *N* means *EJ* or *MJ* multiple electromagnetic transitions to the final $J_f$ state of the nucleus. The numerical coefficient of 31.335 we obtained [7] on the basis of the modern values for fundamental constants [23].

The accurate value of the proton mass [24] and the mass of $^{13}$C equals 13.00335502 amu [25] were used in present calculations. The constant $\hbar^2/m_0$ was taken equal to 41.4686 MeV fm$^2$. The Coulomb potential for $R_{Coul.} = 0$ was written as $V_{Coul.}$(MeV) = $1.439975 \cdot Z_1 Z_2 / r$, where *r* is the distance between the particles in entrance channel in fermi. The Coulomb parameter $\eta = \mu Z_1 Z_2 e^2/(k\hbar^2)$ was presented as $\eta = 3.44476 \cdot 10^{-2} Z_1 Z_2 \mu/k$, where *k* is the wave number in the entrance channel in fm$^{-1}$, determined by the energy *E* of interacting particles $k^2 = 2\mu E/\hbar^2$ [26].

### III. PHASE SHIFT ANALYSIS

While the consideration of the astrophysical *S*-factor of the thermonuclear p$^{13}$C→$^{14}$Nγ radiative capture reaction, the phase shift analysis of the p$^{13}$C elastic scattering at the energies from 250 to 800 keV has been done on the basis of the experimental data obtained in [27,28]. The expressions for the differential cross-sections of the elastic scattering in the system of two particles with spins 1/2 and 1/2 taking into account the spin-orbital splitting of the states and without singlet-triplet phase mixing [29], which can take place in the nuclear systems like N$^3$He, N$^{13}$C etc. [30].

One can see that, as the result of the carried out analysis, the singlet $^1S_0$ phase shift is close to zero (within 1°–3°). Fig. 1 shows the form of the triplet $^3S_1$ phase shift. The triplet $^3S_1$ phase shift has the pronounced resonance corresponding to the level $J^\pi T = 1^-1$ of $^{14}$N in the p$^{13}$C channel at the energy 0.55 MeV (laboratory system) [16]. The width of this resonance has the value 23(1) keV [16] what less than in case of p$^{12}$C scattering [31], and we need for its description the narrow potential without the FS what might lead to the width parameter of the order $\beta = 2 \div 3$ fm$^{-2}$.

The nuclear part of the intercluster potential of the p$^{13}$C interaction is presented as usual; in the Gaussian form [7]

$$V(r) = -V_0 \exp(-\beta r^2)$$

with a point-like Coulomb term given above. The potential for $^3S_1$ wave was constructed so as to describe correctly the resonating partial phase shift (see Fig. 1). Two variants for $^3S_1$ potential of the p$^{13}$C interaction, without FS were obtained using the results of our phase shift analysis:

$$V_S = 265.40 \text{ MeV, and } \beta_S = 3.0 \text{ fm}^{-2}, \tag{2}$$

$$V_S = 186.07 \text{ MeV, and } \beta_S = 2.0 \text{ fm}^{-2}, \tag{3}$$

The calculation results of $^3S_1$ phase shift with such potentials practically coincide and are shown in Fig. 1 by the solid (2) and dashed (3) lines.

The potential with the FS of the $^3P_1$ bound state should represent correctly the bound energy of $^{14}$N with $J^\pi T = 1^+0$ in the p$^{13}$C channel at -7.55063 MeV [16] as well as describe reasonably the mean square radius of $^{14}$N, which has the experimental value 2.560(11) fm [16]. As a result, the following parameters were obtained:

$$V_{GS} = 1277.853205 \text{ MeV and } \beta_{GS} = 1.5 \text{ fm}^{-2}. \tag{4}$$

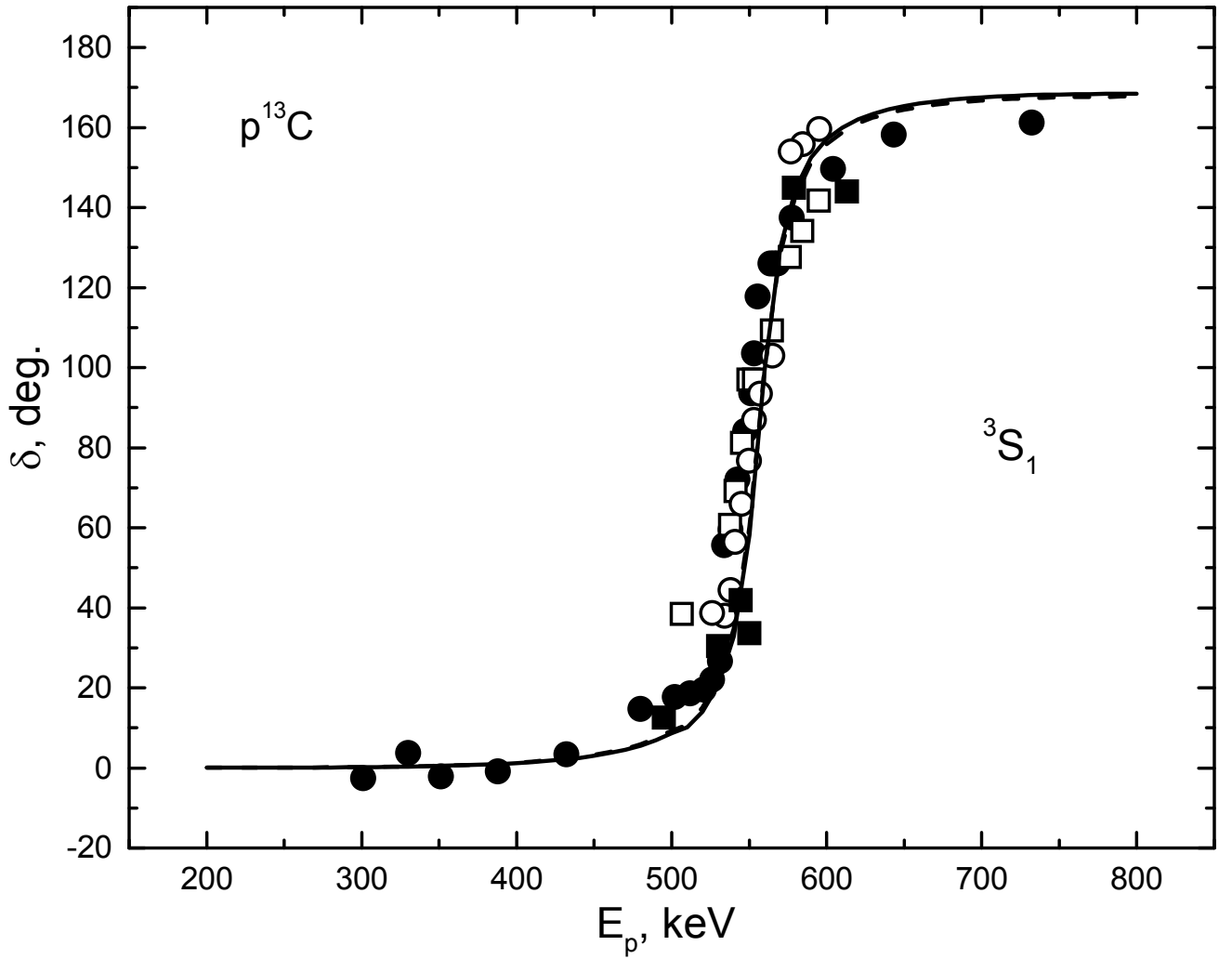

FIG. 1. The $^3S_1$ phase shift of the p$^{13}$C elastic scattering at astrophysical energies. Points: ●, ○, ■ and □ – our phase shift analysis based on the data [27,28]. Lines: solid – the phase shift calculations with potential (2) given in the text, dashed – calculations with potential (3).

The potential gives the bound energy equals -7.550630 MeV and the mean-square radius $R_{ch}$ = 2.38 fm. The values 0.8768(69) fm [24] and 2.4628(39) fm [16] were used as the radii of the proton and $^{13}$C, respectively.

Another variant of the potential the $^3P_1$ ground state of $^{14}$N defined as binary p$^{13}$C system

$$V_{GS} = 1679.445025 \text{ MeV and } \beta_{GS} = 2.0 \text{ fm}^{-2} \quad (5)$$

leads to the bound energy -7.550630 MeV and also gives slightly understated mean-square radius $R_{ch}$ = 2.36 fm.

Note that such a high accuracy of the potential parameters is required for providing of the correct description of nuclear binding energy, up to $10^{-6}$ MeV.

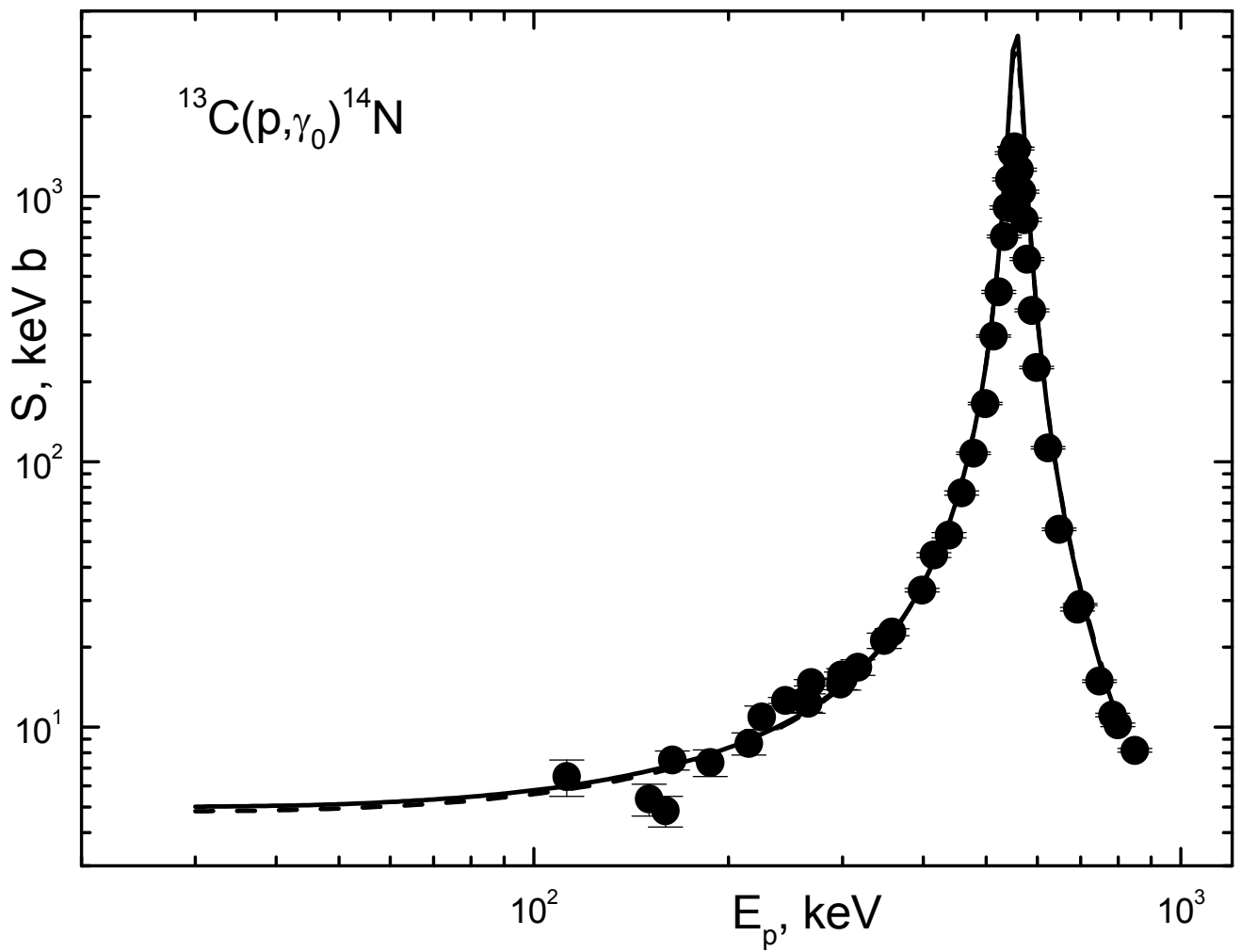

FIG. 2. Astrophysical *S*-factor of the p$^{13}$C radiative capture at low energies. Experimental points (●) are taken from [34]. Lines: solid – our calculation with the first set of potentials (2) and (4), dashed – the calculation with the second set of potentials (3) and (5).

**IV. RADIATIVE CAPTURE**

In this work, we will continue the study of the astrophysical *S*-factors for reactions with light atomic nuclei and will dwell on the p$^{13}$C radiative capture at astrophysical energies. This process is a part of the CNO thermonuclear cycle, which gives the essential contribution into the energy release of thermonuclear reactions [1,6,32,33], leading to the burning of the Sun and the stars of our Universe [11]. The existent experimental data on the astrophysical $S$-factor of the p$^{13}$C radiative capture [21] show the presence of the narrow, with the width about 23(1) keV, resonance at the energy 0.551(1) MeV (laboratory system) [16] that leads to the $S$-factor's rise by two-three orders. Such form of the $S$-factor can be obtained due to the $E$1 transition, with the spin change $\Delta T = 1$, from the $^3S_1$ resonance scattering state at 0.55 MeV and moments $J^{\pi}T = 1^-1$ to the $^3P_1$ triplet bound state of p$^{13}$C clusters, with the potentials like (4) or (5). This state corresponds the ground state of $^{14}$N with quantum numbers $J^{\pi}T = 1^+0$ in the p$^{13}$C channel, since $^{13}$C has the moments $J^{\pi}T = 1/2^-1/2$ [16].

The $S$-factor calculations of the p$^{13}$C radiative capture to the ground state of $^{14}$N with the above-cited potentials, for the $^3P_1$ state and the $^3S_1$ resonance scattering wave at the energies below 0.8 MeV, are given in Fig. 2 by the solid and dashed lines. The experimental results were taken from [34], where, evidently, the most recent investigations of this reaction are reported. In the figure - the solid line is the result of combination of the potentials (2) and (4) and the dashed line for (3) and (5). The calculation results of $S$-factor are very close, since the fact that the phase shifts of two $^3S_1$ scattering potentials, which are shown in Fig. 1, practically coincide. The calculated astrophysical $S$-factor for the first set of potentials (2) and (4) has practically constant value equals 5.0(1) keV b in the energy range 30÷50 keV. The value 4.8(1) keV b was obtained for the second set of potentials (3) and (5) at the same energies. The fixed up error is obtained by the averaging of the $S$-factor value over the above noted energy range.

The known extrapolations of the measured $S$-factor to zero energy for the transitions to the ground state of $^{14}$N lead to the following values: 5.25 keV b [34]; it was obtained 5.16±72 keV b, on the basis of results [34] in [35]; the value 5.36±71 keV b was found in [36]; the value 5.5 keV b was suggested in measurements [37]; and the value 5.06 keV b, for the $S$-factor at zero energy, was given in [38].

### V. CONCLUSION

That is to say, the completely acceptable results in the description of the astrophysical $S$-factor of the p$^{13}$C radiative capture at the energy range from 30 to 800 keV are obtained for both considered variants of p$^{13}$C intercluster potentials. The $S$-factor value is practically constant at the energies 30÷50 keV, thereby determines its value for the extrapolation to zero energy.

The obtained results confirm [7] the fact that to parametrize reliable intercluster scattering potentials and obtain, on their basis, the characteristics of nuclei and nuclear processes is possible only with the enough accurate determination of the experimental data of the elastic cluster scattering phase shifts. The parameters for the BS potentials are fixed by the nuclear characteristics quite definitely.

Unfortunately, at present time, the scattering phase shifts are determined with large errors, for the majority of lightest nuclei; sometimes reach 20-30% [39]. In this connection, in terms of future usage of the PCM, the problem of accuracy increasing for the experimental data of the elastic scattering of light atomic nuclei at low and astrophysical energies, which are needed for the construction of the intercluster potentials, is quite urgent until the present time.